\documentclass[
aps,%
12pt,%
final,%
notitlepage,%
oneside,%
onecolumn,%
nobibnotes,%
nofootinbib,%
superscriptaddress,%
showpacs,%
centertags]%
{revtex4}

\usepackage{amsmath,amssymb,mathrsfs}

\begin{document}

\title{Hamiltonian anomalies of bound states in QED.}

\author{\firstname{V.I.}~\surname{Shilin}}
\affiliation{Joint Institute for Nuclear Research, Dubna, Russia}
\affiliation{Moscow Institute of Physics and Technology, Dolgoprudny, Russia}
\author{\firstname{V.N.}~\surname{Pervushin}}
\affiliation{Joint Institute for Nuclear Research, Dubna, Russia}

\begin{abstract}
The Bound State in QED is described in systematic way by means of nonlocal irreducible representations of the nonhomogeneous Poincare group
and Dirac's method of quantization. As an example of application of this method we calculate triangle
diagram $Para-Positronium \to \gamma\gamma$. We show that the Hamiltonian approach to Bound State in QED leads to anomaly-type contribution to
creation of pair of parapositronium by two photon.
\end{abstract}

\maketitle

\section{Introduction}
The bound states in gauge theories are usually considered in the framework of representations of the homogeneous Lorentz group
in one of the Lorentz-invariant gauges \cite{Hay-91}. In this paper, we suggest a systematic scheme of the bound state generalization
of S-matrix elements, which is based on irreducible representations of the nonhomogeneous Poincar\'e group in concordance with
the first Quantum Electrodynamic (QED) quantization \cite{Dirac,hp} and first QED description of bound states \cite{a15,Salpeter}. We obtain bound states
by means of excluding time-component of four-potential \cite{Dirac,hp,Polubarinov} and Hubbard-Stratonovich transformation \cite{pre-1a,pre-1}, that
is in agreement with general principles \cite{logunov}. The time component is chosen in correspondence
with the Markov-Yukava constraint of irreducibility \cite{MarkovYukawa,kad}.

The aim of the article is to research the experimental consequences of such the Poincar\'e group irreducible representations of QED (see also review \cite{Pervu}).

As a test of this scheme we
calculate process $P \to \gamma\gamma$, where $P$ is parapositronium, that describes triangle diagram, with one real
and the other virtual photons.

The structure of the article is as follows. In section \ref{sec_excluding_time_component} we exclude time-component of four-potential from
QED action. For derived action in section \ref{sec_positronium_classical_action} we make the Hubbard-Stratonovich transformation and
positronium field appear. Than we make semi-classical quantization of the resulting system in section \ref{sec_quantization}. In
section \ref{sec_triangle_anomaly} we calculate the triangle anomaly. And in section \ref{sec_hamiltonian_anomaly} we calculate
the contribution in process $\gamma\gamma \to PP$ inspired by triangle anomaly.

\section{Excluding time component of $A_\mu$ in QED}\label{sec_excluding_time_component}
We start with usual Quantum Electrodynamic Lagrangian:
\begin{equation}
{\cal{L}}_{\rm QED}= -\frac{1}{4}F_{\mu\nu} F^{\mu\nu} - A_{\mu} j^{\mu} -
\overline{\psi}( i\gamma^\mu \partial_\mu - m ) \psi.
\label{lagrangian_qed_initial}
\end{equation}

Below we will work in frame of reference in which the bound state as whole is at rest.

The Lagrangian (\ref{lagrangian_qed_initial}) contains nonphysical degree of freedom. This degree can be excluded by substitution the solution
of classical equation of motion in action.

Variation action providing by (\ref{lagrangian_qed_initial}) over $A_0$ leads to one component of Maxwell equation:
$$
\Delta A_0 - \partial_0 \partial_k A_k = -j_0,
$$
where $\Delta = \partial_1 \partial_1 + \partial_2 \partial_2 + \partial_3 \partial_3$.
If we take the gauge:
\begin{equation}
\partial_k A_k =0
\label{gauge_fixing}
\end{equation}
than the previous equation simplifies:
$$
\Delta A_0 = -j_0.
$$
The solution of this equation is:
\begin{equation}
A_0 = -\frac{1}{\Delta}j_0
\label{solution_Maxwell}
\end{equation}
where
$$
\left(\frac{1}{\Delta}j\right) (x, y, z) = - \frac{1}{4 \pi}
\int\limits_{{\mathbb R^3}\diagdown \{ (x, y, z) \} } dx^\prime dy^\prime dz^\prime
\dfrac{j(x^\prime, y^\prime, z^\prime)}{\sqrt{(x-x^\prime)^2+(y-y^\prime)^2+(z-z^\prime)^2}}
$$

Under gauge fixing condition (\ref{gauge_fixing}) Lagrangian (\ref{lagrangian_qed_initial}) has the form:
$$
{\cal{L}} = \frac{1}{2}\partial_{\mu}A_m \partial^{\mu}A_m-\frac{1}{2}A_0 \Delta A_0-
A_0 j_0+A_m j_m-\overline{\psi}(i\gamma^\mu\partial_\mu-m)\psi
$$
Substitute into this Lagrangian the solution of the classical equation of motion (\ref{solution_Maxwell}):
\begin{equation}
S = \int d^4x \left(\frac{1}{2}(\dot{A_i}\dot{A_i} - B_i B_i)+A_i j_i-
\overline{\psi}(i\gamma^\mu\partial_\mu-m)\psi+\frac{1}{2} j_0\frac{1}{\Delta}j_0 \right)
\label{action_qed_final}
\end{equation}

\section{Positronium effective action}\label{sec_positronium_classical_action}
The quantization of fermion fields in action (\ref{action_qed_final}) yields generating functional:
\addtocounter{equation}{1}{\Large
\begin{equation}
{\mathcal Z} ={\mathscr N}_1 \int {\mathsf D}\bar{\psi} {\mathsf D} \psi e^{\mbox{\normalsize $\displaystyle
iS +
i\int d^4 x \left( \bar{\eta}\psi + \bar{\psi}\eta \right)$}},
\label{generating_functional_initial}
\tag*{\normalsize (\theequation)}
\end{equation}}

$A_i$ here and below we consider only as \emph{external fields}.

In this generating functional consider the term in (\ref{action_qed_final})
$$
\frac{1}{2}\int d^4x \, d^4y \; j_\mu(x) D^{\mu\nu}(x-y) j_\nu(y)=
\int \!\! d^4x \; \frac{1}{2} \: j_0\frac{1}{\Delta}j_0=
$$

\begin{multline}
= -\frac{1}{2} \!\! \int \!\! d^4 \! x_1 d^4 \! x_2 d^3 \! \mathbf{y}_1 d^3 \! \mathbf{y}_2 \;
\bar{\psi}_{\alpha_1}(x_1) \psi^{\alpha_2}(x_2) \\
\underbrace{{\gamma^0}^{\alpha_1}_{\alpha_2} \delta^4(x_1 \!\! - \!\! x_2) \frac{e^2}{4\pi |\mathbf{x}_1 \!\! - \!\! \mathbf{y}_2|}
\delta^3(\mathbf{y}_1 \!\! - \!\! \mathbf{y}_2) {\gamma^0}^{\beta_2}_{\beta_1}
}_{{\cal K}^{\alpha_1 \phantom{\beta_1 \alpha_2} \beta_2}_{\phantom{\alpha_1} \beta_1 \alpha_2}(x_1, \mathbf{y}_1 ; x_2, \mathbf{y}_2)} \\
\bar{\psi}_{\beta_2}(x^0_2,\mathbf{y}_2) \psi^{\beta_1}(x^0_1, \mathbf{y}_1)= \nonumber
\end{multline}

$$
=-\frac{1}{2}\bar{\psi}_{a_1} \psi^{a_2} {\cal K}^{a_1 \phantom{b_1 a_2} b_2}_{\phantom{a_1} b_1 a_2} \bar{\psi}_{b_2} \psi^{b_1} \; = \;
-\frac{1}{2}\bar{\psi}_{a_1} \psi^{b_1} {\cal K}^{a_1 \phantom{b_1 a_2} b_2}_{\phantom{a_1} b_1 a_2} \psi^{a_2} \bar{\psi}_{b_2}
$$
where in the first line $\displaystyle D^{\mu\nu}(x-y) = - \delta^\mu_0 \delta^\nu_0 \frac{e^2 \delta(x^0 - y^0)}{4\pi |\mathbf{x}_1 \!\! - \!\! \mathbf{y}_2|}$, and
in the last line we use next notation: sum over pair indexes $a_1, b_1$ is summing over $\alpha_1, \beta_1$ and integrating
over $d^4 \! x_1 d^3 \! \mathbf{y}_1$.

So, there is forth power of $\psi$ in generating functional. Habbard-Stratonovich transformation allows to reduce power of $\psi$ to second.

Let's consider the combination $\psi \bar{\psi}$ as a real bilocal field:
$$
\chi^{\alpha}_{\phantom{\alpha} \beta}(x^0, \mathbf{x}, \mathbf{y}) \; = \; \psi^{\alpha}(x^0,\mathbf{x}) \bar{\psi}_{\beta}(x^0, \mathbf{y})
$$
Transposition operation we define as:
$$
{\chi^T}_{\alpha}^{\phantom{\alpha} \beta}(x^0, \mathbf{x}, \mathbf{y}) \; = \; -\chi^{\beta}_{\phantom{\beta} \alpha}(x^0, \mathbf{y}, \mathbf{x})
$$
Than introduce new bilocal field ${\mathcal M}^{\alpha}_{\phantom{\alpha} \beta}(x^0, \mathbf{x}, \mathbf{y})$ and make
{\Large
$$
\begin{array}{rcl}
e^{\mbox{\normalsize $\displaystyle i\int \! d^4x \; \frac{1}{2} \: j_0\frac{1}{\Delta}j_0$}} & = &
e^{\mbox{\normalsize $\displaystyle
-\frac{i}{2}\bar{\psi}_{a_1} \psi^{b_1} {\cal K}^{a_1 \phantom{b_1 a_2} b_2}_{\phantom{a_1} b_1 a_2} \psi^{a_2} \bar{\psi}_{b_2}$}}=\\
&=&{\mathscr N} {\displaystyle \int} {\mathsf D} {
\mathcal M} \; e^{\mbox{\normalsize $\displaystyle
\frac{i}{2} {{\mathcal M}^T}_{a_1}^{\phantom{a_1} b_1}
{{\cal K}^{-1}}^{a_1 \phantom{b_1 a_2} b_2}_{\phantom{a_1} b_1 a_2} {\mathcal M}^{a_2}_{\phantom{a_2} b_2} +
i{\bar{\psi}}_a \psi^b {\mathcal M}^a_{\phantom{a} b}$}}
\end{array}
$$}
$\sqrt{\det {\mathcal K}^{-1}}$ is constant and has been included in ${\mathscr N}$.

After that generating functional \ref{generating_functional_initial} takes form
\begin{multline*}
{\mathcal Z} = {\mathscr N}_2 \int 
{\mathsf D}\bar{\psi} {\mathsf D} \psi {\mathsf D}{\mathcal M} \;
\exp{ \: i\left[ \int d^4x \left(\frac{1}{2}(\dot{A_i}\dot{A_i} - B_i B_i)+A_i j_i-
\overline{\psi}(i\gamma^\mu\partial_\mu-m)\psi \right) +\right.}\\
\left. + \frac{1}{2} {{\mathcal M}^T}_{a_1}^{\phantom{a_1} b_1}
{{\cal K}^{-1}}^{a_1 \phantom{b_1 a_2} b_2}_{\phantom{a_1} b_1 a_2} {\mathcal M}^{a_2}_{\phantom{a_2} b_2} +
{\bar{\psi}}_a \psi^b {\mathcal M}^a_{\phantom{a} b} + \int d^4 x \left( \bar{\eta}\psi + \bar{\psi}\eta \right)\right]
\end{multline*}

Introduce notation:
$$
{G_{mA}^{-1}}^{\alpha}_{\phantom{\alpha} \beta}(x,y)\equiv
(-{\mathnormal i}{\gamma^\mu}^{\alpha}_{\phantom{\alpha} \beta} \partial_\mu + m\delta^{\alpha}_{\phantom{\alpha} \beta}
+eA_i {\gamma^i}^{\alpha}_{\phantom{\alpha} \beta})\delta^4(x-y)+
{\mathcal M}^{\alpha}_{\phantom{\alpha} \beta}(x^0, \mathbf{x}, \mathbf{y}) \delta(x^0-y^0)
$$
and define inverse operator as:
$$
\int d^4 \! y {G_{mA}^{-1}}^{\alpha}_{\phantom{\alpha} \beta}(x,y) {G_{mA}}^{\beta}_{\phantom{\beta} \gamma}(y,z) =
\delta^{\alpha}_{\phantom{\alpha} \gamma}\delta^4(x-z)
$$
that allows us to write generating functional in more compact form:
\begin{multline*}
{\mathcal Z} = {\mathscr N}_2 \int {\mathsf D}\bar{\psi} {\mathsf D} \psi {\mathsf D}{\mathcal M} \;
\exp \: i \left[ \int d^4x \left(\frac{1}{2}(\dot{A_i}\dot{A_i} - B_i B_i) \right) +
\frac{1}{2} {{\mathcal M}^T}_{a_1}^{\phantom{a_1} b_1}
{{\cal K}^{-1}}^{a_1 \phantom{b_1 a_2} b_2}_{\phantom{a_1} b_1 a_2} {\mathcal M}^{a_2}_{\phantom{a_2} b_2} + \right.\\
\left. + \int d^4x \, d^4y \; {\bar{\psi}}_\alpha (x) \: {G_{mA}^{-1}}^{\alpha}_{\phantom{\alpha} \beta}(x,y) \: \psi^\beta (y) +
\int d^4 x \left( \bar{\eta}\psi + \bar{\psi}\eta \right)\right]
\end{multline*}

After integration over fermions finally we have:
\begin{multline*}
{\mathcal Z} = {\mathscr N}_3 \int {\mathsf D}{\mathcal M} \;
\exp \: i\left[ \int d^4x \left(\frac{1}{2}(\dot{A_i}\dot{A_i} - B_i B_i) \right) + \frac{1}{2} {{\mathcal M}^T}_{a_1}^{\phantom{a_1} b_1}
{{\cal K}^{-1}}^{a_1 \phantom{b_1 a_2} b_2}_{\phantom{a_1} b_1 a_2} {\mathcal M}^{a_2}_{\phantom{a_2} b_2} - \right. \\
\left. - \int d^4 \! x d^4 \! y {\bar{\eta}}_\alpha (x) {G_{mA}}^{\alpha}_{\phantom{\alpha} \beta}(x,y) \eta^\beta (y) -
i\mathrm{tr}\ln G_{mA}^{-1} \right]
\end{multline*}

So, effective action for positronium field take form:
\begin{equation}
S_P = \int d^4x \left(\frac{1}{2}(\dot{A_i}\dot{A_i} - B_i B_i) \right) +
\frac{1}{2} {{\mathcal M}^T}_{a_1}^{\phantom{a_1} b_1}
{{\cal K}^{-1}}^{a_1 \phantom{b_1 a_2} b_2}_{\phantom{a_1} b_1 a_2} {\mathcal M}^{a_2}_{\phantom{a_2} b_2}
-i\mathrm{tr}\ln G_{mA}^{-1}
\label{action_positronium_initial}
\end{equation}

\section{Quantization of Bilocal Fields}\label{sec_quantization}

For our purpose it would be enough to use the semi-classic approach. According to this approach the first variation of the
action (\ref{action_positronium_initial}) is the Schwinger-Dyson equation :
\begin{equation}
\left. \frac{\delta S_P}{\delta {\mathcal M}} \right|_{A=0}=0
\label{equation_sd_initial}
\end{equation}
that gives a fermion spectrum. And the second variation in the point of minimum is the Bethe-Salpeter equation:
$$
\frac{\delta^2 S_P}{{\delta {\mathcal M}}^2} \Gamma=0
$$
which allows us to find a bound state spectrum.

\subsection{Schwinger-Dyson equation}
Denote solution of (\ref{equation_sd_initial}) as $\Sigma^{\alpha}_{\phantom{\alpha} \beta}(x^0, \mathbf{x}, \mathbf{y}) -
m\delta^{\alpha}_{\phantom{\alpha} \beta}\delta(\mathbf{x}-\mathbf{y})$. Than introduce notation:
$$
{G_{\Sigma}^{-1}}^{\alpha}_{\phantom{\alpha} \beta}(x,y)\equiv
-{\mathnormal i}{\gamma^\mu}^{\alpha}_{\phantom{\alpha} \beta} \partial_\mu +
\Sigma^{\alpha}_{\phantom{\alpha} \beta}(x^0, \mathbf{x}, \mathbf{y}) \delta(x^0-y^0),
$$
where the inverse operator is defined as
$$
\int d^4 \! y {G_{\Sigma}^{-1}}^{\alpha}_{\phantom{\alpha} \beta}(x,y) {G_{\Sigma}}^{\beta}_{\phantom{\beta} \gamma}(y,z) =
\delta^{\alpha}_{\phantom{\alpha} \gamma}\delta^4(x-z).
$$
The equation (\ref{equation_sd_initial}) now takes the form:
\begin{equation}
\Sigma^{a_1}_{\phantom{a_1} b_1} = m \delta^{a_1}_{\phantom{a_1} b_1} +
{\mathnormal i} {\cal K}^{a_1 \phantom{b_1 a_2} b_2}_{\phantom{a_1} b_1 a_2} {G_{\Sigma}}^{a_2}_{\phantom{a_2} b_2},
\label{equation_sd_operator}
\end{equation}
in the last term the sum is over pair indexes $a_2, b_2$ is summing over $\alpha_2, \beta_2$ and integrating
over $d^4 \! x_2 d^4 \! y_2 \delta(x^0_2 - y^0_2)$. Below we also assume a delta-function $\delta(x^0_2 - y^0_2)$ in
contraction of ${\mathcal K}$ and $G$.

After Fourier transform let us takes an ansatz:
$$
\Sigma(p,q) = \delta^4(p \! - \! q)
\left( \hat{\mathbf{q}} \; + \; E(\mathbf{q}) e^{-2\frac{\hat{\mathbf{q}}}{|\mathbf{q}|}\vartheta(\mathbf{q})} \right).
$$
Then we can inverse $G_{\Sigma}^{-1}$:
\begin{equation}
G_{\Sigma}=\left( \frac{\Lambda_+(\mathbf{q})}{q_0-E(\mathbf{q})+{\mathnormal i}\varepsilon} \; + \;
\frac{\Lambda_-(\mathbf{q})}{q_0+E(\mathbf{q})-{\mathnormal i}\varepsilon} \right)\gamma^0,
\label{gsigma}
\end{equation}
where:
\begin{equation}
\Lambda_\pm(\mathbf{q}) \; = \; e^{\frac{\hat{\mathbf{q}}}{|\mathbf{q}|}\vartheta(\mathbf{q})} \frac{1\pm\gamma^0}{2}
e^{-\frac{\hat{\mathbf{q}}}{|\mathbf{q}|}\vartheta(\mathbf{q})}.
\label{lambda}
\end{equation}

Well known that in case of QED Schwinger-Dyson equation gives small correction to electron mass, so we take the solution
of Schwinger-Dyson equation in form $\Sigma^{a}_{\phantom{a} b} \approx m \delta^{a}_{\phantom{a} b}$, where $m$ is the electron mass.

\subsection{Bethe-Salpeter equation}
Taking the second variation of the action (\ref{action_positronium_initial}) in the point of minimum that we have found
solving the Schwinger-Dyson equation (\ref{equation_sd_operator}):
$$
\left. \frac{\delta^2 S_P}{{\delta {\mathcal M}}^2} \right|_{A=0, {\mathcal M}=\Sigma-m} \Gamma=0
$$
we obtain the Bethe-Salpeter equation:
$$
\Gamma^{a_1}_{\phantom{a_1} b_1} = i {\cal K}^{a_1 \phantom{b_1 a_2} b_2}_{\phantom{a_1} b_1 a_2} {G_{\Sigma}}^{a_2}_{\phantom{a_2} a_3}
\Gamma^{a_3}_{\phantom{a_3} b_3} {G_{\Sigma}}^{b_3}_{\phantom{b_3} b_2}.
$$
After Fourier transform and inserting $G_\Sigma$ in (\ref{gsigma}), let us take integral over $q_0$:
\begin{equation}
\Gamma(\mathbf{p}) = \int d^3\!\mathbf{q} 
{\cal K}(\mathbf{p} \! - \! \mathbf{q}) \Psi(\mathbf{q})
\label{equation_bs_gamma_and_wave}
\end{equation}
where the wave-function:
\begin{equation}
\Psi(\mathbf{q}) = 
\gamma^0\left(\frac{\overline{\Lambda}_+(\mathbf{q}) \Gamma(\mathbf{q}) \Lambda_-(\mathbf{q})}{E_P - m_P+{\mathnormal i}\varepsilon} \; + \;
\frac{\overline{\Lambda}_-(\mathbf{q}) \Gamma(\mathbf{q}) \Lambda_+(\mathbf{q})}{E_P + m_P-{\mathnormal i}\varepsilon}\right)\gamma^0
\label{def_wave_function}
\end{equation}
where analogously to (\ref{lambda}) one has:
\begin{equation}
\overline{\Lambda}_\pm(\mathbf{q}) \; = \; e^{-\frac{\hat{\mathbf{q}}}{|\mathbf{q}|}\vartheta(\mathbf{q})} \frac{1\pm\gamma^0}{2}
e^{\frac{\hat{\mathbf{q}}}{|\mathbf{q}|}\vartheta(\mathbf{q})}.
\label{lambdabar}
\end{equation}
Acting (\ref{lambdabar}) and (\ref{lambda}) on (\ref{def_wave_function}), and
inserting (\ref{equation_bs_gamma_and_wave}) in (\ref{def_wave_function}) we have equation for $\Psi$:
$$
(E_P\mp m_P)\Lambda_\pm(\mathbf{p})\Psi(\mathbf{p})\overline{\Lambda}_\mp(\mathbf{p})=
\Lambda_\pm(\mathbf{p}) \left( \int d^3\!\mathbf{q}
{\cal K}(\mathbf{p} \! - \! \mathbf{q}) \Psi(\mathbf{q}) \right)\overline{\Lambda}_\mp(\mathbf{p})
$$
If $|\mathbf{p}| \ll m$ than this equation turns into the Schrodinger one \cite{Pervu}:
\begin{equation}
\left( \frac{1}{2m} \mathbf{p}^2 + (2m-m_P)\right) \Psi(\mathbf{p})=
\int d^3\!\mathbf{q} 
{\cal K}(\mathbf{p} \! - \! \mathbf{q}) \Psi(\mathbf{q})
\label{equation_srod}
\end{equation}

As we interested in parapositronium, we can take the next ansatz:
\begin{equation}
\Psi(\mathbf{q})= N \gamma^5\gamma^0 \psi(\mathbf{p})
\label{ansatz_wave_function}
\end{equation}
where $N$ is unknown normalized factor and will be defined below, $\psi(\mathbf{p})$ is the nonrelativistic wave-function -- the solution
of Schrodinger equation (\ref{equation_srod}): 
\begin{equation}
\psi(r)=\frac{1}{\sqrt{\pi}} \left( \frac{m\alpha}{2} \right)^{\frac{3}{2}} e^{-r\frac{m\alpha}{2}} \qquad \rightarrow \qquad
\psi(\mathbf{p})=\frac{1}{\pi} \frac{(m\alpha)^{\frac{5}{2}}}{{\left({\left(\frac{m\alpha}{2}\right)}^2+\mathbf{p}^2\right)}^2}
\label{formula_shrod_wave_function}
\end{equation}

Because of smallness $\alpha$ the condition $|\mathbf{p}| \ll m$ is satisfied. Therefore the motion of an electron and a positron in the positronium is
nonrelativistic and we can take $M_P \simeq 2m$.

\section{Triangle anomaly}\label{sec_triangle_anomaly}

\begin{figure}[!htb]
\begin{center}
\includegraphics[scale=0.7]{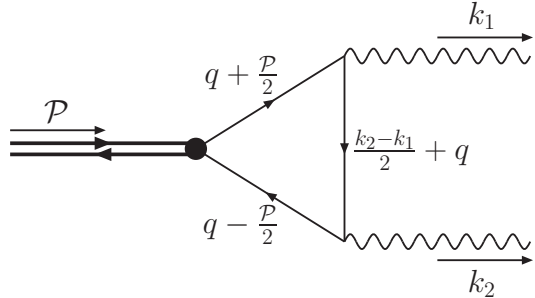}
\caption{Triangle anomaly.}
\label{fig_triangle anomaly}
\end{center}
\end{figure}

Small variation action $S_P$ (\ref{action_positronium_initial}) over ${\mathcal M}$ and double variation over $A$ give us
$$
{\mathcal M}^\prime A_m A_k
\left. \frac{\delta}{\delta {\mathcal M}} \frac{\delta}{\delta A_m} \frac{\delta}{\delta A_k} \right|_{A=0, {\mathcal M}=\Sigma-m} S_P
=-ie^2 A_m A_k \mathrm{tr} \left( \gamma^m G_\Sigma \gamma^k G_\Sigma {\mathcal M}^\prime G_\Sigma \right),
$$
that can be interpreted as the triangle anomaly (fig.\ref{fig_triangle anomaly}) with an amplitude
$$
M_\triangle = e^2 \int d^4\!q \; \mathrm{tr} \! \left[ G_\Sigma(q+\frac{{\cal P}}{2}) \, \Gamma(q) \, G_\Sigma(q-\frac{{\cal P}}{2}) \,
\hat{A}_1(k_1) \, G_\Sigma(\frac{k_2-k_1}{2}+q) \, \hat{A}_2(k_2) \right]
$$

Below in our calculations we consider one virtual photon with momentum $k_1$ and other real photon with momentum $k_2$. For simplicity the reference frame
is chosen as a positronium rest frame ${\mathcal P}=(M_P,0,0,0)$. Let us denote $\mathbf{k}_2$ as $\mathbf{k}$.

Rather crude estimate can be done in the following way. Using (\ref{gsigma}), let us integrate over $q_0$ and take into account only a pole that
forms the wave-function (\ref{def_wave_function}), using (\ref{ansatz_wave_function})
$$
M_\triangle = e^2 N \int \! d^3\!\mathbf{q} \: \psi(\mathbf{q}) \, \mathrm{tr} \! \left( \gamma^5 \gamma^0 \hat{\mathbf{A}}_1(k_1)
\frac{{\displaystyle \frac{\hat{k}_2-\hat{k}_1}{2}}+\hat{q}+m}{2\left( |\mathbf{k}| \sqrt{m^2 + \mathbf{q}^2} - \mathbf{k}\mathbf{q} \right)}
\hat{\mathbf{A}}_2(k_2) \right).
$$
Let us make a change $\gamma^0$ for $\frac{\hat{{\mathcal P}}^0}{M_P}$ and make the trace over the gamma matrixes product
$$
M_\triangle = -4e^2 N \int \! d^3\!\mathbf{q} \: \psi(\mathbf{q}) \varepsilon_{0\lambda\mu\nu} e^\mu_1 e^\nu_2
\frac{{\displaystyle \frac{k^\lambda_2-k^\lambda_1}{2}}+q^\lambda}{2\left( |\mathbf{k}| \sqrt{m^2 + \mathbf{q}^2} - \mathbf{k}\mathbf{q} \right)}
\frac{{\mathcal P}^0}{M_P}.
$$
The term with $q^\lambda$ gives zero contribution. We can change $\varepsilon_{0\lambda\mu\nu} \dfrac{{\mathcal P}^0}{M_P}$ for
sum $\varepsilon_{\rho\lambda\mu\nu} \dfrac{{\mathcal P}^\rho}{M_P}$, furthermore we can suppose now that in arbitrary frame of
reference (not in positronium rest frame of reference as we do before) the last sum should appear. Using (\ref{formula_shrod_wave_function}) one gets:
$$
M_\triangle = 4e^2 N \: \varepsilon_{\mu\nu\lambda\rho} \, k^\mu_1 e^\nu_1 k^\lambda_2 e^\rho_2 \: \frac{1}{2M_P} \frac{(m\alpha)^{\frac{5}{2}}}{\pi}
\int \! d^3\!\mathbf{q} \: \frac{1}{{\left({\left(\frac{m\alpha}{2}\right)}^2+\mathbf{q}^2\right)}^2}
\frac{1}{\left( |\mathbf{k}| \sqrt{m^2 + \mathbf{q}^2} - \mathbf{k}\mathbf{q} \right)}
$$
Notice, that as $\alpha$ is small, so only values of $\mathbf{q}$ near zero are important. Therefore, we can write in spherical
variables, where $\zeta=-\cos\theta$,
\begin{multline*}
2\pi \int\limits^\infty_0 \! dq \: \int\limits^1_{-1} \! dx \: \frac{q^2}{{\left({\left(\frac{m\alpha}{2}\right)}^2+q^2\right)}^2}
\frac{1}{\left( k \sqrt{m^2 + q^2} - kq\zeta \right)} \simeq \\ \simeq
2\pi \left.\int\limits^1_{-1} \! dx \: \frac{1}{\left( k \sqrt{m^2 + q^2} - kq\zeta \right)} \right|_{q \to 0}
\int\limits^\infty_0 \! dq \: \frac{q^2}{{\left({\left(\frac{m\alpha}{2}\right)}^2+q^2\right)}^2}
= 2\pi \frac{2}{km} \frac{\pi}{2\alpha m},
\end{multline*}
and
$$
M_\triangle = \frac{2N\alpha^{\frac{5}{2}}}{\sqrt{m}k} \varepsilon_{\mu\nu\lambda\rho} \, k^\mu_1 e^\nu_1 k^\lambda_2 e^\rho_2.
$$
To get agreement with the decay $P \to \gamma\gamma$ given by anomaly, we should take the normalization factor as
$$
N=\sqrt{2\pi}\sqrt{\frac{m}{2}} \: \frac{1}{2}.
$$

Thus finally we have for the triangle diagram
\begin{equation}
M_\triangle = C_P \varepsilon_{\mu\nu\lambda\rho} \, k^\mu_1 e^\nu_1 k^\lambda_2 e^\rho_2,
\label{triangle_anomaly}
\end{equation}
where
\begin{equation}
C_P = \frac{4m\sqrt{\pi}\alpha^{\frac{5}{2}}}{4m^2-k^2_1}=\frac{2m\sqrt{\pi}\alpha^{\frac{5}{2}}}{(k_2 {\mathcal P})}.
\label{triangle_coeff}
\end{equation}

\section{Hamiltonian anomalies}\label{sec_hamiltonian_anomaly}

In quantization of the Hamiltonian systems we need canonical coordinates and momenta. If the triangle anomaly exists, it changes
the momentum. Consider as an example:
$$
S_{example} = \int d^4x \left(\frac{1}{2}(\dot{A}_i\dot{A}_i - B_i B_i) + C \phi \dot{A}_i B_i \right)
$$
which is the action (\ref{action_positronium_initial}) only with the first term and the triangle
anomaly (\ref{triangle_anomaly}) with ${\mathcal P} = (M_P,0,0,0)$ and $C_P = C$ -- consider as a
constant, $\phi$ is the wave function of positronium.

\begin{figure}[!htb]
\begin{center}
\includegraphics[scale=0.7]{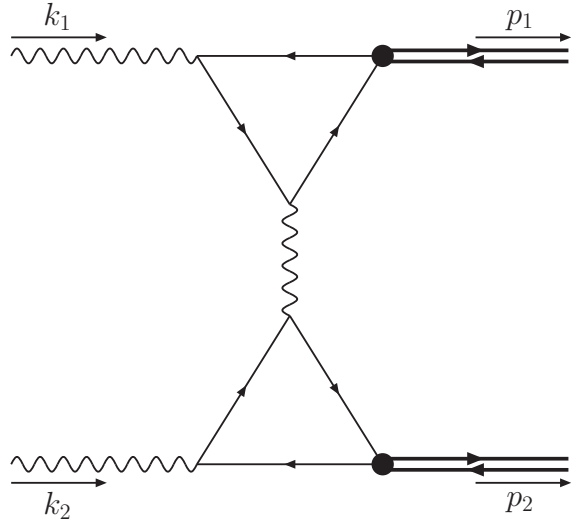}
\caption{Hamiltonian anomaly.}
\label{fig_hamiltonian anomaly}
\end{center}
\end{figure}

The canonicals momentum:
$$
E_i = \frac{\delta S_P}{\delta \dot{A}_i} \neq \dot{A}_i,
$$
so, one can write:
$$
S_{example} = \int d^4 \! x \left( \frac{{(\dot{A}_i+C \phi B_i)}^2 - B_i B_i}{2} \: - \: \frac{1}{2} C^2 \phi^2 B_i B_i\right)
$$
and consider the first term as a "renormalized" electromagnetic field and the second term as a Hamiltonian anomaly shown in fig.\ref{fig_hamiltonian anomaly}.

Back to the action (\ref{action_positronium_initial}) and triangle
anomaly (\ref{triangle_anomaly},\ref{triangle_coeff}), the amplitude of the Hamiltonian anomaly (fig.\ref{fig_hamiltonian anomaly}) is
$$
M_\times = -\frac{1}{2} \frac{C_P(p_1,k_1) C_P(p_2,k_2)}{{M_P}^2}
\varepsilon_{\mu\nu\lambda\rho} p_1^\mu k_1^\lambda e_1^\rho
\varepsilon^{\sigma\nu\xi\zeta} {p_2}_\sigma {k_2}_\xi {e_2}_\zeta.
$$
Evaluating sum over $\nu$
\begin{equation}
M_\times = -\frac{1}{2} \frac{C_P(p_1,k_1) C_P(p_2,k_2)}{{M_P}^2}
\left( V {e_1}^\rho {e_2}_\rho + T^{\mu\nu} {e_1}_\mu {e_2}_\nu \right),
\label{ham_anomaly_amplitude}
\end{equation}
where
\begin{equation}
\begin{array}{rcl}
V &=& -(p_1 p_2)(k_1 k_2) + (p_1 k_2)(k_1 p_2) ,\\
T^{\mu\nu} &=& (p_1 p_2) {k_2}^\mu {k_1}^\nu - (k_1 p_2) {k_2}^\mu {p_1}^\nu - (p_1 k_2) {p_2}^\mu {k_1}^\nu + (k_1 k_2) {p_2}^\mu {p_1}^\nu .
\end{array}
\label{ham_anomaly_amplitude_coeff}
\end{equation}

The square of the amplitude equals to
\begin{multline*}
{\lvert M_\times \rvert}^2 = \frac{\pi^2 \alpha^{10}}{4 {(k_1 p_1)}^2 {(k_2 p_2)}^2} \biggl(
2 (p_1 p_2) (k_1 k_2) (p_1 k_1) (p_2 k_2) + 2 (p_1 k_2) (k_1 p_2) (p_1 k_1) (p_2 k_2) - \\
- 2 {M_P}^2 (p_2 k_1) (p_2 k_2) (k_1 k_2) - 2 {M_P}^2 (k_1 k_2) (p_1 k_1) (p_1 k_2) + {M_P}^4 {(k_1 k_2)}^2 \biggr) .
\end{multline*}

After an integration over the positronium momenta $p_1$ and $p_2$, finally the cross-section is
$$
\sigma = \frac{\pi \alpha^{10}}{96 {M_P}^2 s} \left( (2s + 7{M_P}^2)\sqrt{1- {\left( \frac{2M_P}{s} \right)}^2} -
6{M_P}^2 \ln\frac{\sqrt{s} + \sqrt{s - {(2M_P)}^2}}{\sqrt{s} - \sqrt{s - {(2M_P)}^2}} \right) ,
$$
where $s = {(k_1 + k_2)}^2$ and $M_P$ -- positronium mass.

Consider the behavior of $\sigma$ at small and large $s$ in the center of mass frame of
reference. Denote the energy of one photon as $E_\gamma$, then $s= 4 {E_\gamma}^2$. For $E_\gamma \to M_P$ we have:
$$
\sigma = \frac{\pi \alpha^{10}}{128 {M_P}^2} \sqrt{1- {\left( \frac{M_P}{E_\gamma} \right)}^2} ,
$$
and for $E_\gamma \to \infty$ we have:
$$
\sigma \to \frac{\pi \alpha^{10}}{48 {M_P}^2} .
$$

In conclusion we should notice that in the process $\gamma\gamma \to PP$, where $P$ is the parapositronium
besides (\ref{ham_anomaly_amplitude},\ref{ham_anomaly_amplitude_coeff}) there are leading contributions from the box diagram.

\section{Conclusion}

In this paper we obtain the bound state generation functional by Poincar\'e-invariant irreducible representations. Specific feature of this
method is an absence of explicit relativistic covariance. An attempt to conserve the covariance based on Markov-Yukawa constraint
was considered in \cite{covar}. We will prolong this research in future papers. Some comments about the relativistic covariant excluding
the time component discussed in section \ref{sec_excluding_time_component} can be found in the review \cite{Polubarinov}.

The quantization of the positronium action (\ref{action_positronium_initial}) is made in the semi-classical approach.

As an example of this method, we calculated the triangle anomaly (fig.\ref{fig_triangle anomaly}). To simplify the
calculations the assumption about poles was made. Besides in resulting the answer (\ref{triangle_anomaly},\ref{triangle_coeff}) was
made an assumption about relativistic structure of the answer.

The anomaly triangle diagrams change canonical momenta in Hamiltonian systems. That may leads to additional
contribution (\ref{ham_anomaly_amplitude},\ref{ham_anomaly_amplitude_coeff}) to the
process $\gamma\gamma \to PP$. We can calculate this
contribution using the previous results with the triangle diagram.


\subsection*{Acknowledgments}
The authors would like to thank A.B. Arbuzov, A.Yu. Cherny, A.E. Dorokhov, O.V. Teryaev
and M.K. Volkov, for useful discussions.

\end{document}